\documentclass[preprint,aps,pra,showpacs,floatfix,superscriptaddress]{revtex4-1}
\usepackage{graphicx}
\usepackage{feynmp}
\usepackage{amsmath,amsfonts,amssymb,verbatim,float}

\usepackage[usenames]{color}	
\usepackage{colortbl}
\usepackage{calrsfs}
\usepackage{ulem} 

\usepackage{hyperref} 
\usepackage{subfigure}
\usepackage{indentfirst}
\definecolor{faded}{gray}{0.45}
\newcommand{\bp}{{\bf p}}
\newcommand{\bb}{{\bf b}}
\newcommand{\br}{{\bf r}}
\newcommand{\bbr}{\hat{{\bf r}}}
\newcommand{\bn}{\hat{\bf n}}
\newcommand{\bnu}{\hat{\boldsymbol{\nu}}}

\newcommand{\bsigma}{\boldsymbol{\sigma}}
\graphicspath{{graph/}}
\begin{document}
\thispagestyle{empty}
\title{
Elastic scattering of twisted electrons by atomic target: 
Going beyond the Born approximation
}
\author{V.~P.~Kosheleva}
\affiliation{
Department of Physics, St. Petersburg State University,
Universitetskaya naberezhnaya 7/9, 199034 St. Petersburg, Russia
}
\author{V.~A.~Zaytsev}
\affiliation{
Department of Physics, St. Petersburg State University,
Universitetskaya naberezhnaya 7/9, 199034 St. Petersburg, Russia
}
\author{A.~Surzhykov}
\affiliation{
Physikalisch-Technische Bundesanstalt, 
D-38116 Braunschweig, Germany
}
\affiliation{
Technische Universit\"at Braunschweig,
D-38106 Braunschweig, Germany
}
\author{V.~M.~Shabaev}
\affiliation{
Department of Physics, St. Petersburg State University,
Universitetskaya naberezhnaya 7/9, 199034 St. Petersburg, Russia
}
\author{Th. St\"ohlker}
\affiliation{
GSI Helmholtzzentrum f\"ur Schwerionenforschung GmbH, 
D-64291 Darmstadt, Germany
}
\affiliation{
Helmholtz-Institute Jena, 
D-07743 Jena, Germany
}
\affiliation{
Institut f\"ur Optik und Quantenelektronik, 
Friedrich-Schiller-Universit\"at, D-07743 Jena, Germany
}
%
\begin{abstract}
The elastic scattering of twisted electrons by neutral atoms is 
studied within the fully relativistic framework.
The electron-atom interaction is taken into account in all orders, thus 
allowing us to explore high-order effects beyond the first Born approximation.
In order to illustrate these effects, detailed calculations of the 
total and differential cross sections as well as the degree of polarization
of scattered electrons are performed.
Together with the analysis of the effects beyond the first Born 
approximation, we discuss the influence of the kinematic parameters of 
the incident twisted electrons on the angular and polarization 
properties of the scattered electrons.
\end{abstract}
%
\maketitle
%
%
%
%
%
%
\section{Introduction}
%
%
The twisted (or vortex) electron beams being predicted~\cite{Bliokh} and 
realized~\cite{Verbeeck, Uchida, McMorran} less than a decade ago 
presently attract a lot of attention from both experimental and 
theoretical sides (for a review, see Ref.~\cite{Bliokh_PR_2017}).
In contrast to the conventional (plane-wave) electrons, the twisted 
electrons possess a non-zero projection of the orbital angular momentum 
(OAM) $\hbar m$ onto their propagation direction.
This projection, which nowadays can be as high as $1000\hbar$~\cite{Mafakheri}, 
determines the magnitude of the OAM induced magnetic moment.
Due to such huge (\rm {as opposite to} the plane-wave electrons) magnetic dipole moment 
the twisted electrons are presently regarded as a valuable tool for studying
magnetic properties of materials at nano scale~\cite{Rusz_PRL111_105504:2013, 
Beche_NP10_26:2014, Schattschneider_U136_81:2014, Edstrom_PRL116_127203:2016}.
Moreover, with the growth of $m$ the effects induced by the complex
interplay between the spin and OAM become more pronounced.
All these have stimulated the studies on the interaction of vortex electrons with ionic and atomic targets~\cite{Boxem_PRA89_032715:2014, 
Matula_NJP16_053024:2014, Boxem_PRA91_032703:2015, Serbo_PRA92_012705:2015,
Zaytsev_PRA95_012702:2017}.
%
%
\\
\indent 
%
%
In the present paper, we focus on one of fundamental scattering 
processes, namely, the elastic (Mott) scattering of twisted electrons
by neutral atoms.
This process can be used to better understand the effects induced by a complex interplay 
between the spin and orbital magnetic moments.
These effects can be studied, as an example, by means of measuring magnetic circular 
dichroism in the elastic and inelastic scattering of the vortex 
electron beams~\cite{Rusz_PRL111_105504:2013, Edstrom_PRL116_127203:2016, 
Schattschneider_U179_15:2017, Negi_SR8_4019:2018}, where the spin-orbit 
interaction plays a crucial role.
The case of the Mott scattering by heavy atoms is of particular interest.
Indeed, for such systems the spin-orbit interaction is greatly enhanced.
%
%
Moreover, the Mott scattering is presently regarded as a tool for diagnostics of vortex electrons~\cite{Dutta}, thus stimulating an additional 
interest to the elastic scattering of twisted electrons.
%
%
\\ \indent 
%
%
The Mott scattering of twisted electrons by atoms has been studied in 
Ref.~\cite{Serbo_PRA92_012705:2015} within the first Born approximation.
This approximation is well justified for light atomic systems and relatively 
large velocities of the incident electron.
In the case of the elastic scattering of the twisted electrons by heavy atoms, the description which accounts for the electron-atom 
interaction in all orders is demanded.
We present such description in this paper.
Our approach is based on the non-perturbative treatment of the electron-atom interaction discussed in Ref.~\cite{Zaytsev_PRA95_012702:2017}.
In the framework of this method we evaluate the differential and total 
cross sections for the Mott scattering of the twisted electrons at various 
atomic targets.
The degree of longitudinal polarization for the process considered is also 
calculated.
Additionally, we compare our results with predictions obtained within the first Born approximation~\cite{Serbo_PRA92_012705:2015}.
 It is found that even for light $Z$ systems reliable results for the 
total cross section of the Mott scattering of the twisted electrons with 
energies around (and lower) 100 keV can be obtained only within the all-order 
relativistic treatment.
%
%
\\ \indent 
%
%
The paper is organized as follows.
In Sec.~\ref{sec:pw} the basic equations for the Mott scattering 
of the plane-wave electrons are briefly recalled.
In Sec.~\ref{sec:tw} the theoretical description of the Mott scattering 
of the twisted electrons accounting for the electron-atom interaction in all 
orders is presented.
The comparison of the results obtained within the first Born approximation 
with the all-order relativistic ones is presented in Secs.~\ref{sec:total cross} and~\ref{sec:dcs_deg}. 
Effects of kinematic properties of the incident twisted electron are 
considered in Sec.~\ref{sec:kinematics}.
Finally, a summary and an outlook are given in Sec.~\ref{app:conclusion}. 
%
%
\\ \indent 
%
%
Relativistic units, $m_e = \hbar = c = 1$, and the Heaviside charge 
unit, $e^2 = 4\pi\alpha$ ($\alpha$ is the fine structure constant), are 
used in the paper.
%
%
%
%
%
%
%
\section{BASIC FORMALISM}
\label{sec:basic_form}
The theory of the elastic scattering of the plane-wave electrons by ionic 
and atomic targets is well established and presented in many textbooks
(see, e.g., Ref.~\cite{MottMassey}).
The elastic scattering of the twisted electrons has been considered so 
far only in the framework of the first Born approximation in 
Ref.~\cite{Serbo_PRA92_012705:2015}. 
The approach used in that work can not be utilized for the 
calculations beyond this approximation.
Here we describe the formalism which allows one to study the Mott 
scattering of the twisted electrons without perturbation expansion in 
terms of the coupling constant describing the electron-atom interaction.
This formalism is similar to one appearing in the plane-wave case. 
Let us start, therefore, with the description of the elastic scattering 
of the plane-wave electrons.
%
%
\subsection{Mott scattering of the conventional (plane-wave) electrons}
\label{sec:pw}
The differential cross section (DCS) for the elastic scattering of an 
electron with the asymptotic four-momentum $(\varepsilon, \bp_i)$ and 
the helicity $\mu_i$ (spin projection onto its momentum direction) can be 
expressed as
\begin{equation}
\frac{d\sigma^{(\rm pl)}_{\mu_f\mu_i}}{d\Omega} = 
\left|\tau^{(\rm pl)}_{\mu_f \mu_i}
\right|^{2},
\label{eq:cross}
\end{equation}
where $\mu_f$ is the helicity of the scattered electron possessing four-momentum 
$(\varepsilon, \bp_f)$.
The amplitude $\tau^{(\rm pl)}_{\mu_{f} \mu_{i}}$ of the scattering process
can be written as follows~\cite{Akhiezer}
\begin{equation}
\tau^{(\rm pl)}_{\mu_{f} \mu_{i}} = 
\sqrt{\left(2\pi\right)^{3}}
U^{\dagger}_{p_{f}\mu_{f}}(\bn)
G^{(+)}_{\mu_{i}}\left(\bnu, \bn\right).
\label{eq:ampl_pl}
\end{equation}
Here $\bnu = \bp_i/|\bp_i|$, $\bn= \bp_f/|\bp_f|$,
$G_{\mu_i}^{(+)}(\bnu,\bn)$ is the bispinor amplitude whose explicit 
form will be specified below, and $U_{p_{f}\mu_{f}}(\bn)$ is the Dirac 
bispinor~\cite{Landau4}:
\begin{equation}
U_{p_{f} \mu_{f}}(\bn) = 
\frac{1}{\sqrt{2 \varepsilon}}
\left(\begin{aligned}
& \sqrt{\varepsilon + 1}\chi_{1/2\mu_{f}}(\bn)
\\
& \sqrt{\varepsilon - 1} 
\left( \bsigma \cdot \bn \right) 
\chi_{1/2\mu_{f}}(\bn)
\end{aligned}\right),
\label{eq:dirac_spinor}
\end{equation}
where $\bsigma$ is the vector of Pauli matrices, 
$\chi_{1/2\mu_f}(\bn)$ satisfies $({\bf S}\cdot\bn)\chi_{1/2\mu_f}(\bn) 
= \mu_f \chi_{1/2\mu_f}(\bn)$ with ${\bf S}=\bsigma/2$ being the spin operator. 
%
%
\\
\indent
%
%
The bispinor amplitude  $G_{\mu_i}^{(+)}(\bnu,\bn)$ can be obtained from 
the asymptotic behavior of the incident electron wave function
\begin{equation}
\Psi^{(+)}_{\bp_i\mu_i}\left(\br\right) 
\xrightarrow[r\rightarrow\infty]{} 
\psi_{\bp_i\mu_i}\left(\br\right) 
+ 
G_{\mu_i}^{(+)}(\bnu, \bbr)
\frac{e^{ipr}}{r},
\label{eq:wave_func_as}
\end{equation}
where $p \equiv |\bp_i| = |\bp_f|$, $\bbr$ is the unit vector in the $\br$ direction, 
and $\psi_{\bp_i\mu_i}\left(\br\right)$ is the plane-wave solution of 
the free Dirac equation:
\begin{equation}
\psi_{\bp_i\mu_i}\left(\br\right) = 
\frac{e^{i\bp_i \cdot \br}}
{\sqrt{ \left(2\pi\right)^{3}}}U_{\bp_i \mu_i}(\bnu).
\label{eq:plane_helicity}
\end{equation}
The wave function $\Psi^{(+)}_{\bp_i\mu_i}\left(\br\right)$ describing 
the scattering of the electron in the central potential is given 
by~\cite{Rose,Pratt,Eichler}: 
\begin{equation}
\Psi^{(+)}_{\bp_i\mu_i}(\br) = 
\frac{1}{\sqrt{4 \pi \varepsilon p}} 
\sum_{\kappa m_j} 
C^{j\mu_i}_{l0\ 1/2\mu_i} 
i^l 
\sqrt{2l + 1}
e^{i\delta_{j,l}}
D^{j}_{m_j\mu_i}(\varphi_{\bnu}, \theta_{\bnu}, 0)
\Psi_{\varepsilon\kappa m_j} (\br).
\label{eq:plane_ass}
\end{equation}
Here $\kappa = (-1)^{l + j + 1/2}(j + 1/2)$ is the Dirac quantum number 
determined by the angular momentum $j$ and the parity $l$, 
$C^{JM}_{j_1m_1\ j_2m_2}$ is the Clebsh-Gordan coefficient, $\delta_{j,l}$ 
is the phase shift induced by the scattering potential, $D_{MM'}^{J}$ is 
the Wigner matrix~\cite{Rose2, Varsh}, angles $\varphi_{\bnu}$ and $\theta_{\bnu}$ 
describe the incident electron, and $\Psi_{\varepsilon\kappa m_j} (\br)$ 
is the partial-wave solution of the Dirac equation in the scattering potential $V(r)$ whose explicit form will be specified below. 
Making use of Eqs.~\eqref{eq:wave_func_as},~\eqref{eq:plane_helicity}, 
and~\eqref{eq:plane_ass} one can obtain the explicit expression for the 
bispinor amplitude $G_{\mu_i}^{(+)}(\bnu,\bn)$:
\begin{equation}
G_{\mu_i}^{(+)}(\bnu,\bn)   = 
\frac{1}{\sqrt{2 \varepsilon (2\pi)^{3}}}
\left( \begin{aligned}
& \sqrt{\varepsilon + 1} 
\\
& \sqrt{\varepsilon - 1} \left(\bsigma \cdot \bn\right) 
\end{aligned}\right)
\left[
A + 2B \left(\hat{\boldsymbol{\eta}} \cdot {\bf S} \right)
\right]
\chi_{1/2\mu_i}(\bnu),
\label{eq:G_common_notations}
\end{equation}
where $\hat{\boldsymbol{\eta}} = \left[ \bnu \times \bn \right] / \vert 
\left[ \bnu \times \bn \right] \vert$ and the scattering amplitudes 
$A$ and $B$ express as~\cite{Landau4}:
\begin{eqnarray}
A \equiv A(\bnu\cdot\bn)  & = & \frac{1}{2 i p} 
\sum_{l = 0}^{\infty}
\left\lbrace
\left(l + 1\right)\left[\exp(2i\delta_{l+1/2,l}) - 1 \right]
+ l\left[\exp(2i\delta_{l-1/2,l}) - 1 \right]
\right\rbrace P_{l}\left(\bnu\cdot\bn\right),
\label{eq:amplitude_A}
\\
B \equiv B(\bnu\cdot\bn) & = & \frac{1}{2 p} 
\sum_{l = 1}^{\infty}
\left[
\exp(2i\delta_{l+1/2,l}) - \exp(2i\delta_{l-1/2,l})
\right]
P_{l}^{1}\left(\bnu\cdot\bn\right).
\label{eq:amplitude_B}
\end{eqnarray}
Here $P_{l}$ and $P^{1}_{l}$ are the Legendre polynomials and associated
Legendre functions, respectively.
Substituting Eqs.~\eqref{eq:dirac_spinor} and~\eqref{eq:G_common_notations}
into Eq.~\eqref{eq:ampl_pl} one can express the amplitude of the process
in terms of the amplitudes $A$ and $B$:
\begin{equation}
\tau^{(\rm pl)}_{\mu_f \mu_i} = \chi_{1/2\mu_f}^{\dagger}(\bn)\left[
A + 2B \left(\hat{\boldsymbol{\eta}} \cdot {\bf S} \right)
\right]  \chi_{1/2\mu_i}(\bnu)
\label{eq:ampl}
\end{equation}
and, as a consequence, completely describe the process of the elastic 
electron scattering.
As an example, the DCS for the unpolarized incident electron beam 
expresses as
\begin{equation}
\frac{d\sigma^{(\rm pl)}}{d\Omega} =
\dfrac{1}{2} \sum_{\mu_{f}\mu_{i}} \left|\tau^{(\rm pl)}_{\mu_f \mu_i}
\right|^{2} =
\left|A\right|^{2} + \left|B\right|^{2}. 
\label{eq:cross_plane}
\end{equation}
%
%
\subsection{Mott scattering of the twisted electrons}
\label{sec:tw}
\begin{figure}[H]
\begin{center}
\includegraphics[width=0.65\linewidth]{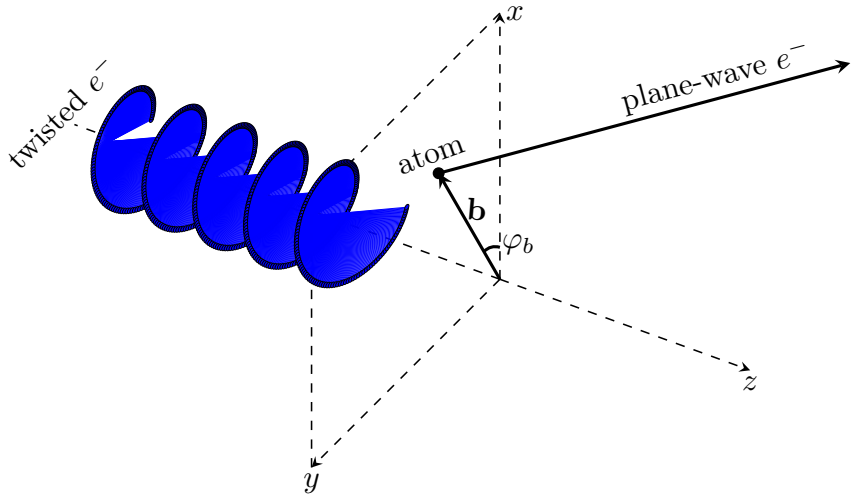}
\caption{ 
The geometry of the Mott scattering of the 
twisted electron on the atomic target being shifted from the $z$ axis on 
the impact parameter $\bb$.
}
\label{ris:mott_scattering}
\end{center}
\end{figure}

%
%
Above we have briefly recalled the theory of the elastic scattering of the
conventional (plane-wave) electrons.
Now we proceed to the consideration of the Mott scattering of the twisted 
electrons.
%
%
\\ \indent
%
%
Let us start with the short theoretical description of twisted electrons.
In the present study we restrict ourselves to the 
case of so-called Bessel electrons.
Free twisted electrons in such a form are the solutions of the Dirac 
equation in an empty space with the well-defined energy $\varepsilon$, 
the helicity $\mu$, and the projections of the linear momentum $p_{z}$ and the 
total angular momentum $m$ onto the propagation direction. Here the $z$-axis is fixed along this direction. 
In addition, the absolute value of the transverse momentum $\varkappa = 
(\varepsilon^{2} - 1 - p_{z}^{2})^{\frac{1}{2}}$ is well defined.
The explicit form of the wave function of the Bessel twisted electron is
given by~\cite{Serbo_PRA92_012705:2015}:
\begin{eqnarray}
\psi_{\varkappa p_{z}m\mu}(\br) & = &
\int 
\frac{e^{im\varphi_{p}} }{2\pi p_{\perp}} 
\delta(p_{\parallel} - p_{z})
\delta(p_{\perp} - \varkappa)
 i^{\mu - m}
\psi_{\bp\mu}(\br)
d\bp,
\label{eq:func_twisted}
\end{eqnarray}
where $p_{\parallel}$ and $p_{\perp}$ are the longitudinal and transversal 
components of the momentum $\bp$, respectively, and $\psi_{\bp\mu}$ is 
the wave function of the plane-wave electron~\eqref{eq:plane_helicity}.
From the form of the integrand in Eq.~\eqref{eq:func_twisted} it is seen 
that the vortex electron can be viewed as a coherent superposition of the
plane-wave electrons with the momenta $\bp$ uniformly distributed
over the surface of a cone with the fixed opening angle $\theta_{p} = 
\arctan(\varkappa/p_{z})$.
%
%
\\ \indent
%
%
Having presented the theoretical description of the free vortex electrons, 
we now turn to the Mott scattering of these electrons by a single atom. 
To begin with, we fix the geometry of the considered process as shown 
in Fig.~\ref{ris:mott_scattering}.
\noindent
In this figure the Mott scattering of the twisted electron by a single 
atom being shifted from the $z$ axis on the impact parameter $\bb = (\rm b_{x}, \rm b_{y}, 0)$ is 
depicted.
It should be noted that the 
relative position of the target and the incident vortex electron is  
important.
This is explained by the fact that the intensity profile and the flow of 
the twisted electron are not homogeneous functions of the space 
variables~\cite{Serbo_PRA92_012705:2015, Zaytsev_PRA95_012702:2017}. 
Additionally, here and throughout we consider that the scattered electron 
is asymptotically described as the plane-wave one.
This is done in accordance with the assumption that in the experiments the 
detectors of the plane-wave electrons are utilized. 
%
%
\\ \indent
%
%
Let us now proceed to the theoretical description of the Mott scattering of
the twisted electrons.
This description can be naturally performed via the construction 
of the amplitude of the process considered.
In order to construct the amplitude incorporating the electron-atom 
interaction nonperturbatively, one can utilize the wave function of the 
twisted electron being a solution of the Dirac equation in the external 
field of the corresponding atomic target.
The explicit form of such a wave function for the process depicted in 
Fig.~\ref{ris:mott_scattering} is~\cite{Zaytsev_PRA95_012702:2017}
\begin{equation}
\Psi^{(+)}_{\varkappa m p_{z} \mu_i}(\br + \bb ) = 
\int 
\frac{e^{im\varphi_{p}} }{2\pi p_{\perp}} 
\delta(p_{z} - p_{\parallel})
\delta(p_{\perp} - \varkappa)
 i^{\mu_{i} - m}
e^{i\bp \cdot \bb}
\Psi^{(+)}_{\bp\mu_i}(\br)
d\bp.
\label{eq:exact_w_f}
\end{equation}
The subsequent construction of the amplitude of the Mott scattering is 
performed in a way similar to one presented in Section~\ref{sec:pw}. 
In the framework of this approach, the amplitude for the scattering of Bessel electrons
depicted in Fig.~\ref{ris:mott_scattering} can be written as
\begin{equation}
\tau^{(\rm tw)}_{\mu_{f} \mu_{i}} = 
\sqrt{\left(2\pi\right)^{3}}
U^{\dagger}_{\bp_f \mu_f }(\bn)
G^{(+)}_{m\mu_i,\bb}(\theta_p,\bn),
\label{eq:ampl}
\end{equation}
where $G^{(+)}_{m\mu_i,\bb}(\theta_p,\bn)$ is the bispinor amplitude 
which is defined as follows 
\begin{equation}
\Psi^{(+)}_{\varkappa m p_{z} \mu_i}(\br + \bb)
\xrightarrow[r\rightarrow\infty]{}
\psi_{\varkappa m p_{z} \mu_i}(\br + \bb) +
G^{(+)}_{m\mu_i,\bb}(\theta_p,\bbr) \frac{e^{ipr}}{r}.
\label{eq:wf_tw_asymp}
\end{equation}
From Eq.~\eqref{eq:wf_tw_asymp} one can derive 
\begin{equation}
G^{(+)}_{m \mu_i, \bb}(\theta_{p}, \bbr) = \int d\bp 
\frac{e^{im\varphi_{p}} }{2\pi p_{\perp}} 
\delta(p_{\parallel} - p_{z})
\delta(p_{\perp} - \varkappa)
 i^{\mu_{i} - m} e^{i\bp \cdot \bb} 
G_{\mu_i}^{(+)}(\hat{\bp},{ \bbr}),
\label{eq:tw_bispinor} 
\end{equation}
where $\hat{\bp}$ is the unit vector in the $\bp$ direction and $G_{\mu_i}^{(+)}(\hat{\bp}, \bbr)$ is the bispinor amplitude of the 
elastic scattering of the plane-wave electrons~\eqref{eq:G_common_notations}.
Inserting Eq.~\eqref{eq:tw_bispinor} into Eq.~\eqref{eq:ampl}, one 
obtains the amplitude for the Mott scattering of the twisted electron by
the single atom
\begin{equation}
\tau^{(\rm tw)}_{\mu_{f} \mu_{i}}(\bb) = 
 \int
d\bp
\frac{e^{im\varphi_{p}} }{2\pi p_{\perp}} 
\delta(p_{z} - p_{\parallel})
\delta(p_{\perp} - \varkappa)
 i^{\mu_{i} - m}
e^{i\bp \cdot \bb}
\tau_{\mu_{f} \mu_{i}}^{\rm (pl)}.
\label{eq:tay_twist}
\end{equation}
Since this amplitude contains the complete information about the process 
under investigation we regard the theoretical description of the Mott
scattering of the twisted electrons as complete.
%
%
%
%
\\ \indent
%
%
Until now we have discussed the elastic scattering of the twisted 
electron by a single atom. 
This process is interesting from theoretical viewpoint but it can hardly 
be realized in experiment.
We consider, therefore, a more realistic scenario in which the twisted 
electron beam collides with a macroscopic target, which we describe as 
an incoherent superposition of atoms being randomly and homogeneously distributed.
The differential cross section for this case is given by~%
\cite{Serbo_PRA92_012705:2015}:
\begin{equation}
\frac{d\sigma^{(\rm tw)}_{\mu_{f} \mu_{i}}}{d\Omega} = 
\int \frac{d\bb}{\pi R^2} \left\vert \tau^{(\rm tw)}_{\mu_{f} \mu_{i}}(\bb) \right\vert ^2
=
\frac{1}{\cos\theta_{p}}
\int \frac{d\varphi_{p}}{2\pi}
\frac{d\sigma^{(\rm pl)}_{\mu_{f} \mu_{i}}}{d\Omega},
\label{eq:exact_cross}
\end{equation}
where $1/(\pi R^2)$ is the cross section area with $R$ being the radius 
of the cylindrical box.
From Eq.~\eqref{eq:exact_cross} it is seen that in the case of the elastic 
scattering of the twisted electrons by the macroscopic target the DCS does not depend on the TAM projection.
%
%
%
%
%
%
%
%
\section{RESULTS AND DISCUSSIONS}
\label{app:results}
Here we restrict our consideration to the Mott scattering of the twisted electrons 
by the target consisting of the neutral atoms. 
Following work~\cite{Serbo_PRA92_012705:2015}, we describe the 
electrostatic potential of a neutral atom as was suggested in Refs.~\cite{Salvat, Salvat_1987}, 
namely, by a sum of three Yukawa terms
\begin{equation}
V(r) = -\dfrac{\alpha Z}{r}\sum_{i=1}^{3}A_{i}e^{-\alpha_i r},
\label{eq:potential}
\end{equation}
where $A_{i}$ is the amplitude of the potential and $\alpha_i$ is the 
scaling constant. 
In the framework of the Born approximation, this potential allows one 
to obtain the analytical expression for the Mott scattering amplitude~\cite{Serbo_PRA92_012705:2015}.
In order to obtain the results beyond the Born approximation, one needs
to evaluate the amplitudes~\eqref{eq:amplitude_A} and~\eqref{eq:amplitude_B}
which determine uniquely the amplitudes~\eqref{eq:ampl} 
and~\eqref{eq:tay_twist}.
This requires the knowledge of the phase shifts $\delta_{j,l}$ 
being induced by the scattering potential~\eqref{eq:potential}.
Here these phases are numerically found with the usage of the 
modified RADIAL package~\cite{Salvat1995}.
As an independent check, the phase shifts $\delta_{j,l}$ were found by means
of the variable phase method~\cite{Calogero, Babikov, Grechukhin_JETPL60_779:1994}.
%
%
\subsection{Total scattering cross sections}
\label{sec:total cross}
\begin{figure}[H]
\begin{center}
\includegraphics[width=0.75\linewidth]{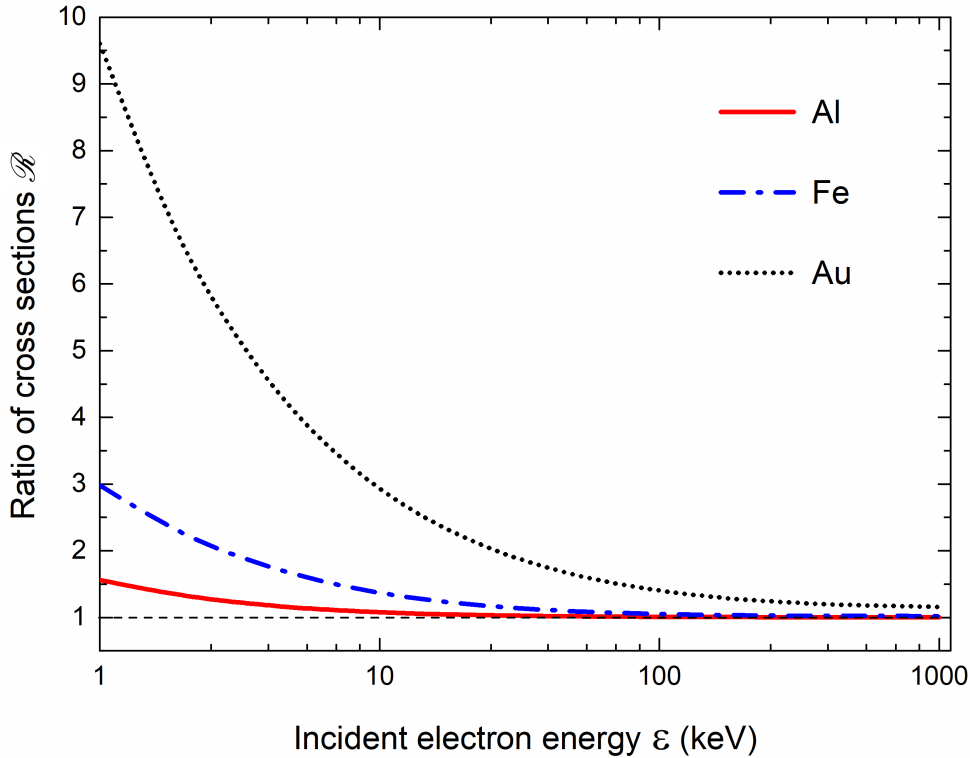}
\caption{
The $\mathcal{R}$ parameter~\eqref{eq:R} as a function of the incident electron energy.
The solid (red), dash-dotted (blue), and dotted (black) lines correspond to the scattering by the aluminium ($Z = 13$), iron ($Z = 26$), 
and gold ($Z = 79$) mactroscopic targets, respectively.
}
\label{ris:magn}
\end{center}
\end{figure}

%
%
Let us compare the total cross sections obtained within the Born, 
$\sigma^{\rm (tw,\ B)}_{\rm tot}$, and all-order, $\sigma_{\rm tot}^{(\rm tw)}$, 
approaches.
For this purpose, we introduce the ratio of the cross sections:
\begin{equation}
\mathcal{R} = \frac{\sigma_{\rm tot}^{\rm (tw,\ B)}}
         {\sigma_{\rm tot}^{\rm (tw)    }}.
         \label{eq:R}
\end{equation}
The explicit expression for $\sigma^{\rm (tw,\ B)}_{\rm tot}$ can be 
found in Ref.~\cite{Serbo_PRA92_012705:2015}.
The all-order relativistic total cross section, $\sigma^{(\rm tw)}_{\rm tot}$, is 
obtained from Eq.~\eqref{eq:exact_cross} via integration over the solid angle 
of the scattered electron and performing the summation and 
averaging over the final- and initial-state helicities, respectively.
Here it is worth stressing that the $\mathcal{R}$ parameter does not depend on the 
opening (conical) angle of the vortex electron $\theta_p$.
Indeed, from Eq.~\eqref{eq:exact_cross} one can deduce that the total 
cross sections for the Mott scattering of the twisted and plane-wave 
electrons are connected as follows

\begin{equation}
\sigma_{\rm tot}^{(\rm tw)} 
=
\frac{\sigma_{\rm tot}^{(\rm pl)}}{\cos\theta_p}.
\label{eq:sigma_tot}
\end{equation}
From this equation one can conclude that the $\mathcal{R}$
parameter coincides with the same parameter for the plane-wave electrons.
\\
\indent
In Fig.~\ref{ris:magn} the ratio~\eqref{eq:R} is presented as a 
function of the incident electron energy.
 It should be noted that this ratio
can be also obtained from the results of 
Ref.~\cite{Salvat} where the Mott scattering of the plane-wave electrons 
has been studied.
The values presented in Fig.~\ref{ris:magn} are in excellent agreement 
with ones from that paper, what supports our numerical calculations.
\noindent
From this figure one can see that for the scattering by such light atoms as aluminium ($Z = 13$) the results obtained within the Born 
approximation agree with the all-order relativistic ones on the level better than 10\% 
already at energies higher than 10 keV.
In the case of the elastic scattering by iron ($Z = 26$) atoms which was 
studied in Ref.~\cite{Serbo_PRA92_012705:2015} this accuracy can be 
achieved only at energies higher than 100 keV and amounts to few 
percents for 1 MeV.
This means that the reliable value of the total cross section for the 
Mott scattering of the twisted electrons with energies around (and lower)
100 keV can be obtained only within the all-order relativistic treatment.
For the Mott scattering by the gold ($Z = 79$) target, meanwhile, the 
difference between the results obtained in the Born approximation and 
beyond is larger than 15\% at 1 MeV and reaches 40\% at energies 
around 100 keV.
One can conclude, therefore, that the Born approximation can not be 
applied for such heavy systems as gold ($Z = 79$) for 
energies up to 1 MeV. 
%
%
%
\subsection{Differential cross sections and degree of polarization}
\label{sec:dcs_deg}
\begin{figure}[H]
\begin{center}
\includegraphics[width=0.75\linewidth]{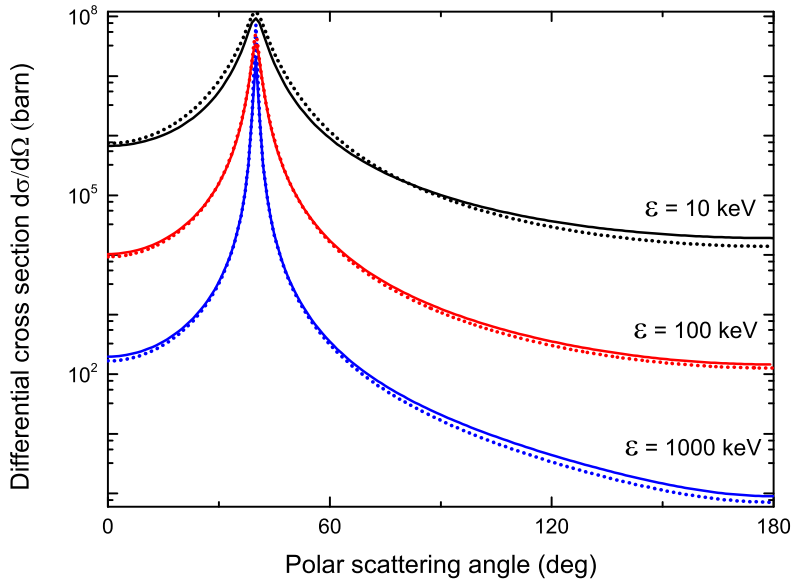}
\caption{
The differential cross section for the Mott scattering of the twisted 
electrons with the opening (conical) angle $\theta_p = 40^{\circ}$ by the 
macroscopic iron ($Z = 26$) target. 
Results of the Born approximation (dotted lines) are compared with 
predictions of the all-order relativistic approach (solid lines).
The kinetic energies of the incident electron are 10 keV, 100 keV, and 1 MeV.
}
\label{ris:DCS_FE}
\end{center}
\end{figure}

\begin{figure}[H]
\begin{center}
\includegraphics[width=1\linewidth]{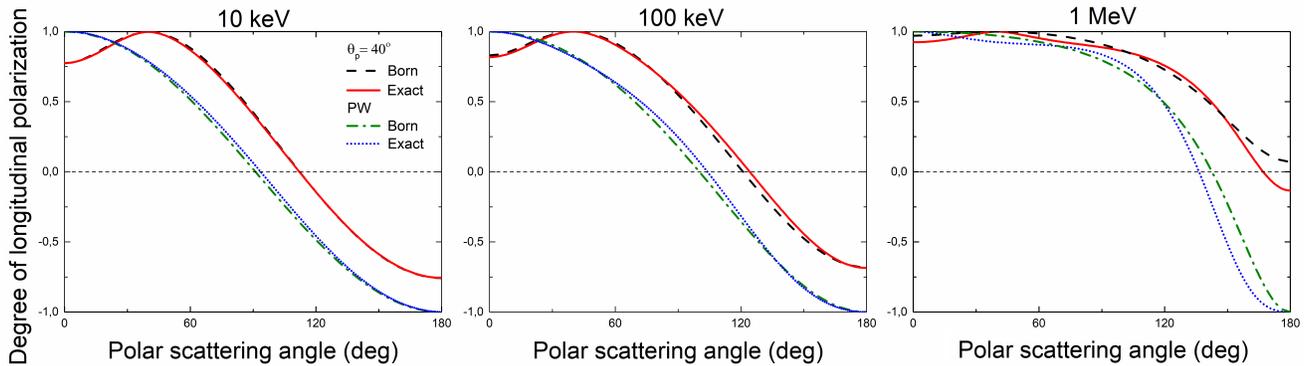}
\caption{
The degree of the longitudinal polarization $P$, defined by Eq.~\eqref{eq:deg_pol}, for 
the Mott scattering of the twisted and plane-wave electrons
by the macroscopic iron ($Z = 26$) target. 
Calculations within the all-order approach and Born approximation have been 
performed for the kinetic energy $\varepsilon = 10$ keV (left panel), 
100 keV (middle panel) and 1 MeV (right panel).
}
\label{ris:deg_26}
\end{center}
\end{figure}

%
%
Let us now compare the differential cross sections calculated 
within the Born and all-order relativistic approach.
 To this end, in Fig.~\ref{ris:DCS_FE} we present the DCS for the Mott 
scattering of the twisted electron with the opening (conical) angle 
$\theta_p = 40^\circ$ by the macroscopic iron ($Z = 26$) target.
\noindent
From this figure it is seen that the qualitative behavior of the DCS
can be well described in the framework of the Born approximation~%
\cite{Serbo_PRA92_012705:2015}. Indeed, as in the Born approximation, the DCS for the Mott 
scattering of the twisted electrons has a peak at polar scattering angle 
$\theta = \theta_p$. 
However, it is worth stressing that the absolute value of the DCS 
obtained in this approximation can strongly differ from the all-order 
relativistic result.
For instance, in the case of the Mott scattering of the 100 keV twisted 
electron the difference between the DCS obtained within the Born approximation 
and the all-order approach reaches $16\%$ for the scattering angle $120^\circ$. 
%
%
\\
\indent
%
%
So far we have compared the total and differential cross sections obtained 
within the Born approximation and beyond it.
For both these measurables the necessity of the all-order relativistic calculations was 
found.
From our point of view, it is also important to investigate the effects beyond the Born approximation for the relative 
measurable quantities. 
In the framework of the present investigation, we consider
one of them, namely, the degree of longitudinal polarization of the 
scattered electrons $P$:
\begin{equation}
P = \dfrac{ d\sigma_{1/2,\ 1/2} - d\sigma_{1/2,\ -1/2} }
          { d\sigma_{1/2,\ 1/2} + d\sigma_{1/2,\ -1/2} },
\label{eq:deg_pol}
\end{equation}
where $d\sigma_{\mu_f\mu_i} \equiv \frac{d\sigma_{\mu_f\mu_i}}{d\Omega}$ 
and it is assumed 
that the incident electron is completely longitudinally polarized 
($\mu_i = 1/2$).
In Fig.~\ref{ris:deg_26} we present the degree of the longitudinal 
polarization $P$ for the Mott scattering of the twisted electron by the 
macroscopic iron ($Z = 26$) target.
\noindent
From this figure one can see that for the kinetic energies of the incident 
electron 10 keV and 100 keV (left and middle panels, respectively)
the degree of polarization obtained within the Born approximation almost 
coincide with the all-order result.
The most interesting situation occurs for the Mott scattering of the 1 MeV
electron (right panel in Fig.~\ref{ris:deg_26}).
For this energy one might expect a better agreement of the results obtained 
within the Born approximation and the all-order relativistic treatment.
However, our calculations show that the corresponding difference 
of the degree of polarization is the most pronounced for the Mott 
scattering of the 1 MeV electron.
From our point of view, it can be explained by the enhancement of the 
spin effects with the increase of the incident electron energy which are 
not accounted in the framework of the Born approximation~\cite{Landau4}.
%
%
\subsection{Effects of kinematic properties of twisted electrons}
\label{sec:kinematics}
\begin{figure}
\centering
\includegraphics[width=0.75\linewidth]{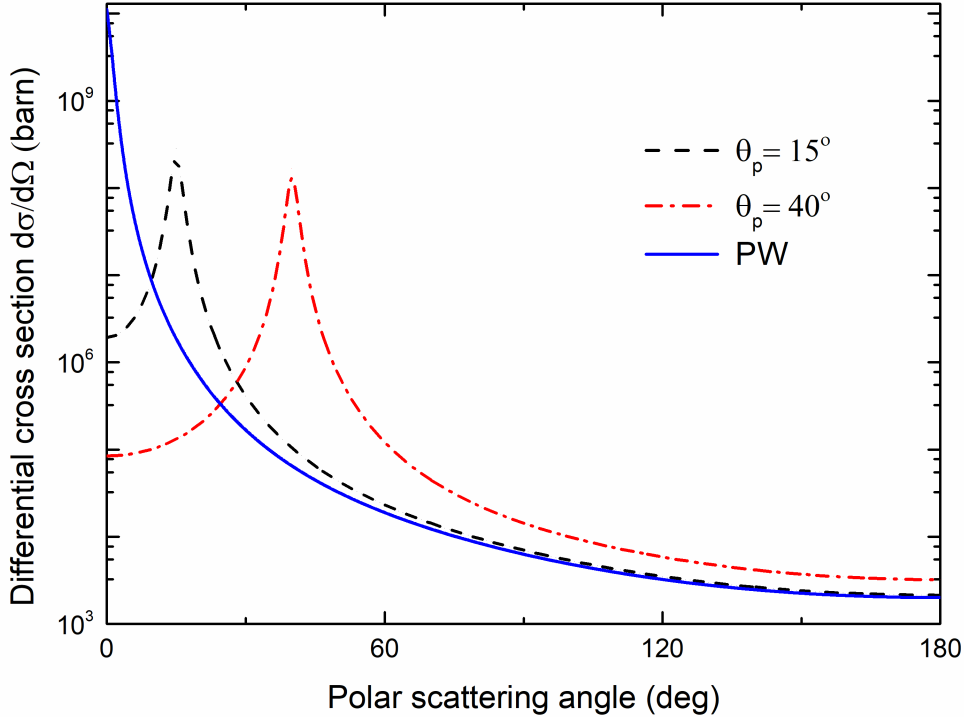}
\caption{
The differential cross sections for the Mott scattering of 100 keV electrons by the 
macroscopic gold ($Z = 79$) target. 
The solid (blue) line corresponds to the plane-wave case, meanwhile the 
dashed (black) and dash-dotted (red) lines are related to the scattering
of the twisted electron with the opening angles $\theta_p = 15^\circ$
and $40^\circ$, respectively.
}
\label{ris:neitral}
\end{figure}

%
%
We finish this section by investigating the dependence of the angular and 
polarization properties of the Mott scattering on the kinematic 
parameters of the incident twisted electron.
In particular, the sensitivity of the differential cross section and the 
degree of the longitudinal polarization to variations of the energy and 
the opening (conical) angle of the vortex electron is studied.
%
%
 In the present section, we investigate the dependence from the kinematic 
properties of the incident twisted electrons for such a heavy element as gold 
($Z = 79$).
Additional interest to this system is explained by the fact that 
the golden foil is usually used in Mott detectors~\cite{Mott, Kessler}.
As shown in Section~\ref{sec:total cross}, the description of this process 
can not be performed within the Born approximation.
Below we, therefore, present only the results of the all-order calculations.
%
%
\\ \indent
%
%
%
%
The DCS for the Mott scattering by the macroscopic gold target is 
presented in Fig.~\ref{ris:neitral}.
\noindent
From this figure it is seen that the DCS for the Mott scattering by the 
golden foil does not differ qualitatively from the DCS for the elastic 
scattering by other macroscopic targets (see Section~\ref{sec:total cross} and 
Ref.~\cite{Serbo_PRA92_012705:2015}).
Namely, the electrons are predominantly scattered under the angle $\theta = \theta_p$. 
And, although the qualitative behavior of the DCS for the scattering by
the macroscopic gold target can be well described in the framework of 
the Born approximation, we would like to stress that the results obtained 
within this approximation can not be regarded as reliable.
\begin{figure}
\begin{center}
\includegraphics[width=\linewidth]{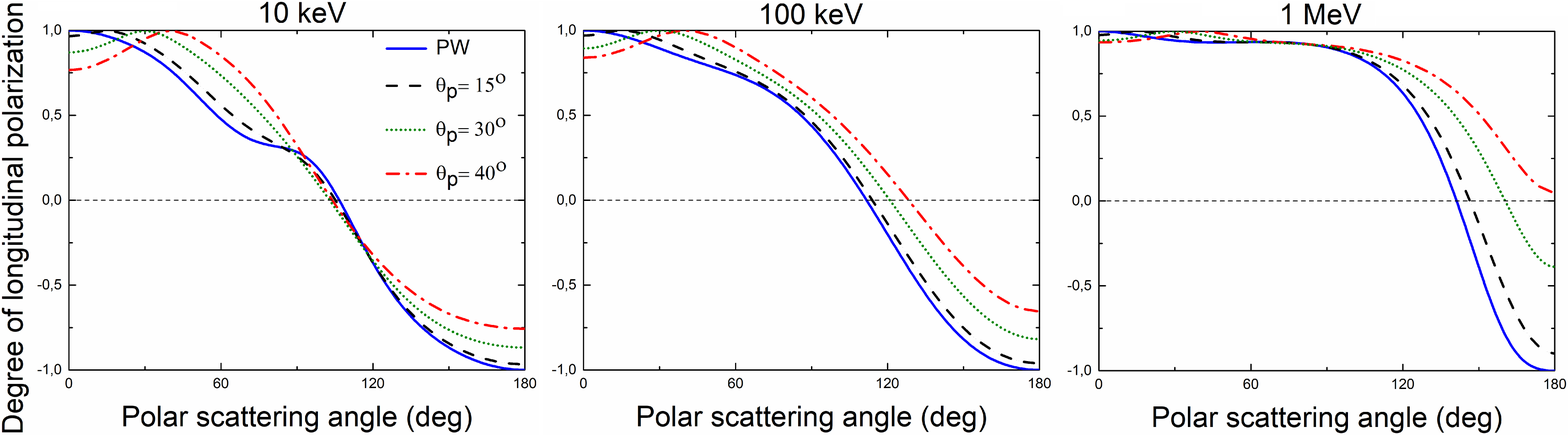}
\caption{
The degree of polarization $P$, defined by 
Eq.~\eqref{eq:deg_pol}, for the Mott scattering by the macroscopic gold 
($Z = 79$) target. 
The solid (blue) line corresponds to the plane-wave case, meanwhile the 
dashed (black), dotted (green) and dash-dotted (red) lines are related to the scattering
of the twisted electron with the opening angles $\theta_p = 15^\circ$, 
$ \theta_p = 30^\circ$, and $ \theta_p = 40^\circ$, respectively.
The left, middle, and right panels correspond to the cases of 10 keV, 
100 keV, and 1 MeV kinetic energies of the incident electron, respectively.
}
\label{ris:deg_pol_Au}
\end{center}
\end{figure}
%
%
\indent
%
%
The degree of the polarization $P$ for the Mott scattering 
by the macroscopic gold target is depicted in Fig.~\ref{ris:deg_pol_Au}.
This figure demonstrates the strong dependence of $P$ from the opening 
angle $\theta_p$ for all energies of the incident electron.
Moreover, this dependence becomes the most prominent for large energies 
(see the right panel in Fig.~\ref{ris:deg_pol_Au}). 
From this panel it is also seen that at certain conical angles the degree of 
the longitudinal polarization is strictly positive.
%
%
%
%
%
%
%
%
\section{CONCLUSION}
\label{app:conclusion}
In the present work, we have considered the Mott scattering of the twisted 
electrons by the atomic targets taking into account the electron-atom 
interaction in all orders.
For the nonperturbative treatment of this interaction the method based 
on the results of Ref.~\cite{Zaytsev_PRA95_012702:2017} was employed.
In the framework of this method, the differential and total cross 
sections and the degree of longitudinal polarization for the Mott 
scattering of the twisted electrons by various macroscopic targets
were evaluated.
%
%
\\
\indent 
%
%
The comparison of the results being nonperturbative in electron-atom 
interaction with ones obtained within the Born approximation was performed. 
It was found that the reliable value of the total cross section for the
Mott scattering of the twisted electrons with energies around (and lower) 
100 keV by the iron ($Z = 26$) target can be obtained only within the all-order relativistic treatment.
In the case of the scattering by the gold ($Z = 79$) target it was found
that the Born approximation can not be applied for energies lower than 1 MeV.
{In contrast, the} qualitative behaviour of the differential cross section and the 
degree of longitudinal polarization can be well described 
in this approximation.
%
%
\\
\indent 
%
%
Additionally, the results obtained within the all-order relativistic 
treatment were presented for the elastic scattering of the twisted electrons at
the golden foil of the Mott detector. 
For such target the strong dependence of the degree of the longitudinal 
polarization from the opening angle $\theta_p$ was found.
%
%
%
%
%
%
%
%
\section*{ACKNOWLEDGEMENTS}
%
%
This work was supported %
by the grant of the President of the Russian Federation (Grant No. MK-4468.2018.2),
by RFBR (Grants No. 18-32-00602 and No. 16-02-00334),
by the German-Russian Interdisciplinary Science Center (G-RISC) funded by the German Federal
Foreign Office via the German Academic Exchange Service (DAAD), and
by SPbSU-DFG (Grants No. 11.65.41.2017 and No. STO 346/5-1).
%
%
%
%
%

%
\end{document}